\documentclass[prl, preprint, superscriptaddress]{revtex4}

\usepackage{graphicx}% Include figure files
\usepackage{dcolumn}% Align table columns on decimal point
\usepackage{bm}% bold math

\begin{document}

\title{Delta Doping of Ferromagnetism in Antiferromagnetic Manganite Superlattices} 

\author{T.S. Santos}
 \affiliation{Center for Nanoscale Materials, Argonne National Laboratory, Argonne, IL 60439}
\author{B.J. Kirby}
 \affiliation{NIST Center for Neutron Research, NIST, Gaithersburg, MD 20899} 
\author{S. Kumar}
 \affiliation{Institute for Theoretical Solid State Physics, IFW Dresden, 01171 Dresden, Germany}
\author{S.J. May}
 \affiliation{Department of Materials Science and Engineering, Drexel University, Philadelphia, PA}
 \affiliation{Materials Science Division, Argonne National Laboratory, Argonne, IL 60439}
\author{J.A. Borchers}
 \affiliation{NIST Center for Neutron Research, NIST, Gaithersburg, MD 20899} 
\author{B.B. Maranville}
 \affiliation{NIST Center for Neutron Research, NIST, Gaithersburg, MD 20899} 
\author{J. Zarestky}
 \affiliation{Ames Laboratory and Department of Physics and Astronomy, Iowa State University, Ames, IA 50011}
\author{S. G. E. te Velthuis}
\affiliation{Materials Science Division, Argonne National Laboratory, Argonne, IL 60439}
\author{J. van den Brink}
 \affiliation{Institute for Theoretical Solid State Physics, IFW Dresden, 01171 Dresden, Germany}
\author{A. Bhattacharya}
 \affiliation{Center for Nanoscale Materials, Argonne National Laboratory, Argonne, IL 60439}
 \affiliation{Materials Science Division, Argonne National Laboratory, Argonne, IL 60439}

\begin{abstract}
We demonstrate that delta-doping can be used to create a dimensionally confined region of metallic ferromagnetism in an antiferromagnetic (AF) manganite host, without introducing any explicit disorder due to dopants or frustration of spins. Delta-doped carriers are inserted into a manganite superlattice (SL) by a digital-synthesis technique. Theoretical consideration of these additional carriers show that they cause a local enhancement of ferromagnetic (F) double-exchange with respect to AF superexchange, resulting in local canting of the AF spins.  This leads to a highly modulated magnetization, as measured by polarized neutron reflectometry. The spatial modulation of the canting is related to the spreading of charge from the doped layer, and establishes a fundamental length scale for charge transfer, transformation of orbital occupancy and magnetic order in these manganites. Furthermore, we confirm the existence of the canted, AF state as was predicted by de Gennes [P.-G.\ de Gennes, Phys.\ Rev.\ {\bf 118} 141 1960], but had remained elusive.
\end{abstract}
\maketitle

In semiconductors, delta-doping \cite{Dingle} has led to record high mobilities in two-dimensional electron gases, spawning a number of fundamental discoveries and many novel applications. In the complex oxides, similar doping strategies may in principle be used to create two-dimensional analogs of the many collective phases found in these materials \cite{Bozovic, Kozuka}. In this work, using AF La$_{1-x}$Sr$_x$MnO$_3$ as the host material, we have devised a delta-doping strategy for locally tailoring the strength of the F double-exchange interactions relative to the AF superexchange, creating a quasi two-dimensional region of enhanced moment. We do this without introducing a random disorder potential due to dopants, or the frustration inherent to some AF/F interfaces.  Our approach is unique to systems where double-exchange (DE) and superexchange (SE) interactions compete, and cannot be realized in conventional semiconductors or metals.

In the seminal work of de Gennes \cite{deGennes}, it was recognized that when the insulating AF parent compound of a manganite is doped with carriers, competition between DE and SE causes the AF spins to cant, in proportion to the doping.  In real materials, additional effects such as the Jahn-Teller (JT) interactions, on-site Coulomb repulsions (Mott-Hubbard effects) and orbital ordering instabilities, all favor localizing the carriers and compete with DE.  As a result, the transition between an AF insulator such as LaMnO$_3$ to a F metal such as La$_{0.80}$Sr$_{0.20}$MnO$_3$ does not follow de Gennes' phase diagram but rather takes place through a mixed-phase region for intermediate values of $x$ \cite{Kawano}. A tendency toward phase separation is believed to be a fundamental property of the double-exchange manganites \cite{Khomskii} and may thwart the appearance of a homogeneous, canted AF phase \cite{Dagotto}.

The situation is quite different if one starts with La$_{1-x}$Sr$_{x}$MnO$_3$ for $x>0.5$, which is also AF. The Mn $e_g$ electrons  can go into one of two partially filled bands made of degenerate $d_{3z^2-r^2}$ and $d_{x^2-y^2}$ orbitals. For $0.5 < x < 0.7$ this degeneracy is removed by the formation of a highly anisotropic $A$-type AF order (planes of F spins that are mutually AF), with the electrons mainly occupying the $d_{x^2-y^2}$ orbital states \cite{Brink}.  The transport is nearly metallic in-plane due to DE.  However, the out-of-plane resistivity is orders of magnitude higher since the $d_{3z^2-r^2}$ states are unoccupied and SE dominates in this direction \cite{Kuwahara}. Upon doping with electrons toward $x<0.5$, the $d_{3z^2-r^2}$ orbitals begin to be occupied, DE begins to act in the out-of-plane direction, and the material transforms into a 3-dimensional F metal.  We show in this work that by delta-doping an AF manganite SL with electrons, we locally enhance the DE which cants the spins away from AF alignment, resulting in a highly modulated magnetization. Our theoretical investigation shows that a JT distortion is instrumental in stabilizing this canted spin structure lying in the midst of the crossover from a 2D antiferromagnet to a 3D ferromagnet near $x=0.5$.

We synthesized cation-ordered analogs of La$_{0.5}$Sr$_{0.5}$MnO$_3$ by alternating single unit cell layers of LaMnO$_3$ (LMO) and SrMnO$_3$ (SMO) with atomic layer precision onto SrTiO$_3$ substrates using ozone-assisted molecular beam epitaxy.  Neutron diffraction measurements on these (LaMnO$_3)_1$/(SrMnO$_3)_1$ superlattices confirmed the $A$-type AF phase, and magnetometry measurements using a superconducting quantum interference device (SQUID) showed a net F moment near zero at low field \cite{Santos_AFmetal}.  Here, we present delta-doped SLs with nominal compositions $x=0.44$ and 0.47. The $x=0.44$ SL was made by alternating unit cell layers of LaMnO$_3$ and SrMnO$_3$ and inserting an additional LaMnO$_3$ layer for every four LaMnO$_3$/SrMnO$_3$ bilayers \cite{suppl_mat}.  This sequence was repeated 9 times (9 supercells), to form the superlattice [(SMO1/LMO1)x4,LMO1]x9.  A La$_{0.56}$Sr$_{0.44}$MnO$_3$ alloy film with a thickness of 81~uc was made for direct comparison.  The $x=0.47$ SL was made by inserting an extra LaMnO$_3$ layer after every 9 LaMnO$_3$/SrMnO$_3$ bilayers, repeated 4 times: [(SMO1/LMO1)x4,LMO1,(SMO1/LMO1)x5]x4. Structural characterization of the SLs is discussed elsewhere \cite{suppl_mat, Santos_AFmetal}. By periodically inserting the extra LaMnO$_3$ unit cell layer, we have delta-doped the MnO$_2$ planes in the vicinity of the inserted LaMnO$_3$ unit cell with electrons. Our investigation of these SLs by neutron scattering techniques was aimed at determining the influence of these delta-doped charges on the $A$-type spin structure.

\begin{figure}
\includegraphics[width=8.5cm]{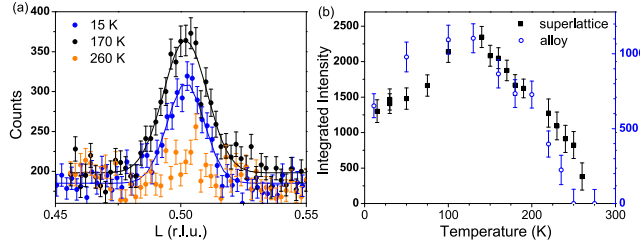}
\caption{\label{fig:diffraction} (a) (0 0 $\frac{1}{2}$) neutron diffraction peak for the $x=0.44$ SL. The lines are a guide to the eye. (b) $T$-dependence of the integrated intensity of the diffraction peak for the $x=0.44$ SL and alloy, measured while increasing $T$ in zero field, after cooling in zero field.}
\end{figure}

We performed neutron diffraction measurements to probe the spin structure in the $x=0.44$ SL and alloy.  The $A$-type order is manifested by a structurally-forbidden (0 0 $\frac{1}{2}$) Bragg peak, shown in Fig.~\ref{fig:diffraction}a, signifying an AF spin alignment with a periodicity of $2c$ normal to the film plane ($c$ is the out-of-plane lattice parameter).  From the full width at half-maximum of the peak, a magnetic coherence length of $\sim$28.5~nm was determined, nearly the entire film thickness (30.9~nm). The coexistence of both a (0 0 $\frac{1}{2}$) neutron diffraction peak and a net F moment from $M(H)$ data \cite{suppl_mat} suggests a \textit{canted} AF structure, for which the spins within a single plane are ferromagnetically aligned and the spins on adjacent planes are canted at a relative angle $\theta_z<180^{\circ}$, where $\theta_z/2$ is the angle between each spin and the applied magnetic field direction. The temperature $(T)$ dependence of the peak intensity in Fig.~\ref{fig:diffraction}b indicates that the $A$-type order in the SL disappears at $\sim$260~K. 

\begin{figure}
\includegraphics[width=8.5cm]{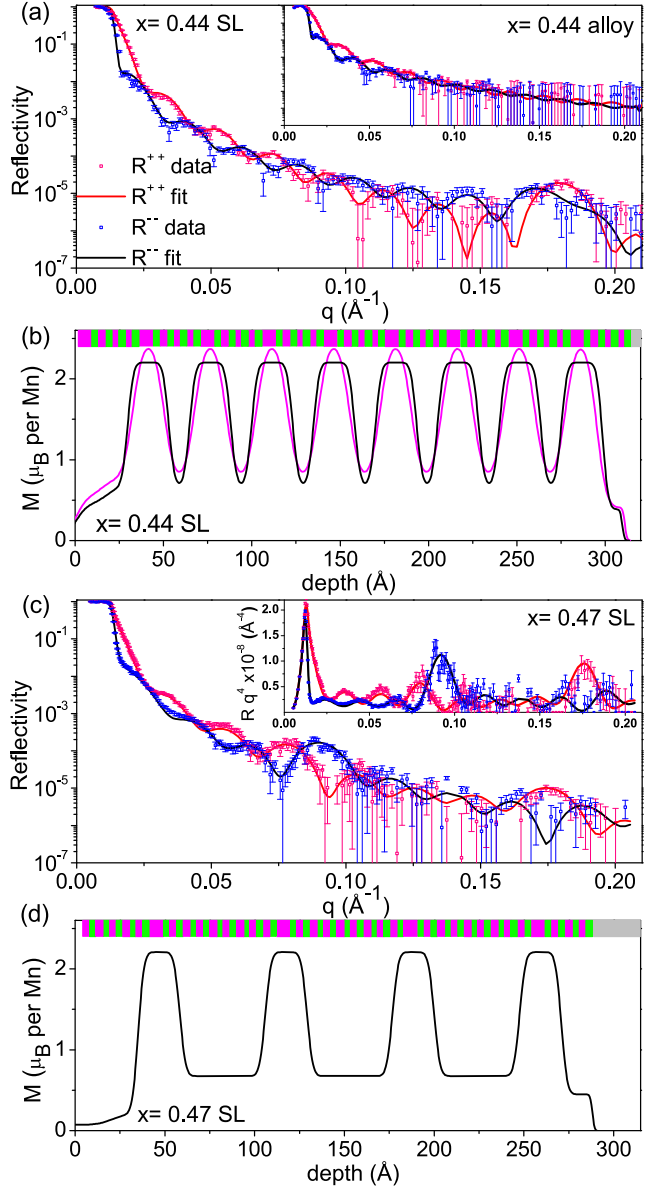}
\caption{\label{fig:profile} (a) The non-spin-flip channels ($R^{++}$ and $R^{- -}$) of the PNR for the $x=0.44$ SL at $T = 120$~K and with $H = 675$~mT applied in the plane of the film, has a Bragg peak signifying that the magnetization is modulated. The solid lines are the fit to the data. The inset shows the PNR of the $x=0.44$ alloy film having no Bragg peak. (b) Two possible magnetic profiles as a function of film depth from the fit to the PNR spectra of the $x=0.44$ SL. The location of the LMO (pink) and SMO (green) layers in the SL and the SrTiO$_3$ substrate (gray) are shown. (c) PNR for the $x=0.47$ SL with $H=815$~mT and $T=120$~K. The Bragg peaks are more apparent in the inset, which shows the same data multiplied by $q^4$ to compensate for Fresnel decay. (d) The magnetic depth profile of the $x=0.47$ SL from the best fit.}
\end{figure}

To determine the depth-dependent magnetization profile, we performed polarized neutron reflectometry (PNR) measurements, which detect variations of in-plane magnetization as a function of sample depth along the surface normal. The non-spin-flip reflectivities, $R^{+ +}$ and $R^{- -}$ shown in Fig.~\ref{fig:profile}a for the $x=0.44$ SL, are sensitive to the component of magnetization parallel to the applied field direction.  Because the nuclear scattering length densities of SMO and LMO are nearly identical, only magnetic variations in the layer profile are detectable.  The key feature of interest here is the Bragg peak at $q=0.18$~\AA$^{-1}$, indicating that the magnetization is in fact \textit{modulated} with a periodicity that matches the supercell periodicity of the SL.

To fit the PNR data using the co\_refine routine \cite{Fitz_chapter}, a model consisting of 9 supercells was used, where each supercell consisted of a high moment sublayer and a low moment sublayer. The sublayer thicknesses, magnetic roughnesses and magnetic scattering length densities (a quantity directly proportional to the magnetization) of the seven interior supercells were fit while constrained to be identical, whereas the topmost and bottommost supercells could vary independently. For this data and model, there are several sets of possible fit parameters that yield the same lowest $\chi^2$ value to within 1\%.  Two such profiles are shown in Fig.~\ref{fig:profile}b. Even though the highest (lowest) moment in the modulation period, $M_{max}$ ($M_{min}$) can each vary by as much as 0.25~$\mu_B$, we can conclude with certainty across all sets of fit parameters that 1) the period of the magnetic modulation matches the structural supercell to within 0.1~nm; 2) the high moment sublayer contains the extra LMO layer and 3) the magnitude of the modulation is quite large, with $M_{max}/M_{min} = 3\pm0.3$.

The $x=0.47$ SL has a thicker supercell with one extra LMO layer for every 19~uc, such that \textit{two} Bragg peaks are detectable, as shown in the PNR spectra in Fig.~\ref{fig:profile}c. With this additional information on the magnetic modulation compared to the $x=0.44$ SL data set, the fitting routine produces a single set of best fit parameters at the lowest $\chi^2$ for the $x=0.47$ SL.  A model similar to the one previously described, containing alternating high moment and low moment sublayers, was used to fit this data.  The moment is again highly modulated, alternating between a high moment region across 6~uc (the full-width half-maximum of the peaks in Fig.~\ref{fig:profile}d) containing the extra LMO unit cell and having $M_{max}=2.2 \mu_B$, and a low moment region spanning 13~uc having $M_{min}=0.7 \mu_B$. The gradual transition between $M_{max}$ and $M_{min}$ occurs over 4~uc, and as we shall show, this occurs by varying the canting angle. Using $M_{tot}=(4-x) \mu_B$ as the magnitude of the moment of a Mn$^{3+/4+}$ spin and $M=M_{tot} cos(\theta_z/2)$, the measured modulation of $M$ corresponds to a modulation of canting angle between $\theta_z=103^{\circ}$ and $157^{\circ}$. These fit parameters also work for the $x=0.44$ SL, where the high moment region extends across 6~uc with $M_{max}=2.2 \mu_B$, and the low moment region extends across 3~uc with $M_{min}=0.7 \mu_B$, as shown in Fig.~\ref{fig:profile}b (black line). In comparison, the PNR measurement of the $x=0.44$ alloy film (inset of Fig.~\ref{fig:profile}a) shows a net moment without any modulation (no Bragg peak). Thus, the spin canting in the alloy is uniform, and the modulated magnetization in the $x=0.44$ and 0.47 SLs occurs due to the delta-doping.

\begin{figure}
\includegraphics[width=8.5cm]{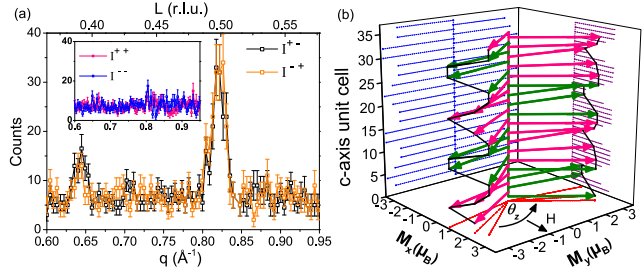}
\caption{\label{fig:spinflip} (a) Diffraction with polarized neutrons and polarization analysis, measured at $T=120$~K in a 820~mT field. Peaks in the spin-flip intensities at high $q$ directly confirm the $canted, modulated$ spin structure. The inset shows the non-spin flip intensities. (b) Schematic of the canted, modulated spin structure, showing 4 supercells (9 uc each) of the $x=0.44$ SL.  Pink and green arrows represent the F spin alignment within the MnO$_2$ planes of the LMO and SMO layers, respectively. The dotted blue (purple) lines are the projection of the spins onto the $yz$-plane ($xz$-plane) and signify the magnetic potential measured by neutron diffraction (PNR). The dotted red lines are the projection of the spins onto the $xy$-plane, showing the canting angles. The black lines aid the eye in identifying the spin modulation.}
\end{figure}

To investigate this canted spin structure further, we carried out diffraction measurements with polarized neutrons and polarization analysis in the vicinity of the AF peak for the $x=0.44$ SL, as shown in Fig.~\ref{fig:spinflip}a.  The peak in spin-flip intensities, $I^{+ -}$ and $I^{- +}$, at $q=2\pi/(2c)=0.82$~\AA$^{-1}$ results from the canted AF spins with each sublattice having a component of magnetization pointing perpendicular to the applied field direction, with a period of $2c$. Moreover, the superlattice peak at $q=0.64$~\AA$^{-1}$ signifies that the canted spin structure is modulated with a period of 9~uc, consistent with the supercell period and the periodicity of the F component revealed by the low $q$ PNR measurements \cite{note_qvalues}.  These peaks in $I^{+ -}$ and $I^{- +}$ directly confirm the \textit{canted, modulated} spin structure. Fig.~\ref{fig:spinflip}b shows the canted, modulated spin structure as deduced by the PNR and diffraction measurements.

In order to explore theoretically this canted spin state, we consider the standard two-orbital model for manganites \cite{Dagotto} with the Hamiltonian $H=H_{DE} + H_{SE} + H_{JT}$ \cite{suppl_mat}. The double-exchange term $H_{DE}$, which arises from a combination of kinetic energy and strong Hund's rule coupling, favors the F state. The superexchange term $H_{SE}$ is the Heisenberg interaction between $t_{2g}$-spins favoring AF coupling. The Jahn-Teller term $H_{JT}$ represents the coupling of the $e_g$ electrons to the lattice distortions.

\begin{figure}
\includegraphics[width=8.5cm]{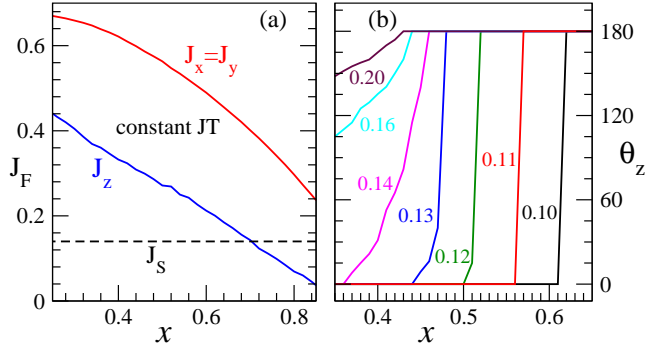}
\caption{\label{fig:theory} (a) The effective F exchange constant arising from the DE interaction, shown here for a \textit{fixed} JT distortion and a \textit{ferromagnetic} spin state for all $x$. The presence of a JT distortion leads to a difference in orbital occupancy, which in turn leads to anisotropic exchange parameters. A typical value of the AF coupling, $J_S=0.14t$ is shown as a dashed line. (b) Doping dependence of the canting angle $\theta_z$ for various values of $J_S$. The canting angle is obtained by minimizing the total energy at a given $x$, where both the JT distortion and $\theta_z$ are allowed to vary.}
\end{figure}

Before understanding the canted magnetic states, it is essential to understand the stability of $A$-type AF order near $x=0.5$ in manganites. It is an outcome of a competition between the ferromagnetic $H_{DE} = J_{F}\sum cos(\theta_{ij}/2)$ and the antiferromagnetic $H_{SE} = J_{S} \sum cos(\theta_{ij})$ terms, where $\theta_{ij}$ denotes the relative angle between spins ${\bf S}_i$ and ${\bf S}_j$, and the summation is over all nearest neighbors. $A$-type order requires the F term to dominate in the $xy$-plane and the AF term to be stronger along the $z$-axis. The presence of a uniform JT distortion favors the occupation of $d_{x^2-y^2}$ orbitals, which leads to a strong F interaction in the $xy$-plane and a much weaker one along the $z$-axis, as seen in Fig.~\ref{fig:theory}a ($J^{z}_{F}\sim p_zt_z$, where $p_z$ is the $d_{3z^2-r^2}$ orbital occupancy and $t_z$ is the hopping integral in the $z$ direction). This leads to $A$-type AF order for $x>0.5$ when the SE coupling $J_S>>J^{z}_{F}$. Upon reducing $x$ so that filling of the $e_g$ band continues, the doped electrons begin to fill the higher $e_g$ states (the $d_{3z^2-r^2}$ states), so that $J^{z}_{F}$ increases and competes with $J_S$ while the in-plane order remains unaffected, as shown in Fig.~\ref{fig:theory}a. Such a competition between F and AF interactions along the $z$-axis can lead to canted spin states. From a simple energy minimization $dE/d\theta_z = 0$ for $E = -J^z_{F} cos(\theta_z/2) + J_S cos(\theta_z)$, one finds that the canting angle $\theta_z = 2 \arccos(J^z_{F}/4J_S)$ if $J^z_{F} < 4J_S$, and $\theta_z = 0$ (F alignment) when $J^z_{F} > 4J_S$. Note that within this scenario one needs $J_{S} = \infty$ to obtain $\theta_z = 180^{\circ}$ (AF state). In reality there is a feedback effect of the spin states on the orbital occupancies that causes the variation of $\theta_z$ to be more abrupt. When the spins cant away from F alignment, this causes occupancy of the $d_{3z^2-r^2}$ orbitals to decrease, which in turn causes $J^z_{F}$ to decrease further, making the spins cant even more. The cost in elastic energy in creating the lattice distortion with changes in orbital occupancy ($c$-axis shrinks with lower $p_z$) prevents the transition from being a step function.

In order to test this explicitly, we minimize the total energy of the Hamiltonian for variational parameters $\theta_z$ \textit{and} JT distortions, hence allowing for this feedback effect.  At a fixed $x$, JT distortions lead to a reduction in electronic energy but there is also an elastic cost associated with making these distortions \cite{suppl_mat}. The optimum value of the distortion is selected from the above competition. This generates an $x$ dependence on the magnitude of the distortions and therefore an additional $x$ dependence on the values of $J^{z}_{F}$. The resulting optimum angle $\theta_z$ is plotted in Fig.~\ref{fig:theory}b for a few values of $J_S$. The sharp change in $\theta_z$ from $180^{\circ}$ to $0^{\circ}$ for $J_S=0.10$, 0.11 in the calculation does not allow for layers at intermediate $x$ with $0^{\circ}<\theta_z<180^{\circ}$. This implies that it is not possible to stabilize a finite canting angle for these parameters. However, for $J_S\geq0.12$ and for $x<0.5$, the variation in $\theta_z$ with $x$ is more gradual and layers with a finite canting angle are stable. Therefore, a small variation in the doping level can lead to significant changes in canting angle and thus the net F moment. The large modulation in moment as a function of depth in these delta-doped SLs is a realization of this scenario. Attempts to realize the canted AF state near $x=0$ as proposed by de Gennes and near $x=0.5$ have typically been obscured by phase segregation into F and AF domains \cite{Chmaissem}, which is mitigated in our digital synthesis approach of charge doping without disorder. We note that a modulation in orbital occupancy is expected to accompany this modulated spin state \cite{Kiyama}.

In conclusion, we have employed delta-doping to tailor the magnetic structure following the ideas of de Gennes. Localized regions of enhanced F exchange were created in an otherwise AF structure, where the transition between the high moment region and the low moment region occurs via a continuous variation of canting angle that is governed by the spreading of charge near the delta-doped layer.  Furthermore, the PNR technique enables us to determine a length scale of 6~uc for the region of enhanced moment that is induced by the delta-doped layer. The result that the SL magnetization is highly modulated is evidence that the doped charges are not completely delocalized over the entire SL. Charge spreading out to 3~uc from both sides of the delta-doped layer is consistent with previous theoretical and experimental works that had inferred a length scale for charge spreading via more indirect means \cite{Millis, May_PRB07}.  We believe this to signify a fundamental length scale for the spreading of charge normal to the layers.

\begin{acknowledgments}
We are grateful to Chuck Majkrzak for helpful suggestions. Work at Argonne and use of the Center for Nanoscale Materials was supported by the U. S. Department of Energy, Office of Science, Office of Basic Energy Sciences, under Contract No. DE-AC02-06CH11357.  A portion of this research at Oak Ridge National Laboratory's High Flux Isotope Reactor was sponsored by the Scientific User Facilities Division, Office of Basic Energy Sciences, U.S. Department of Energy.  We acknowledge the support of the National Institute of Standards and Technology, U.S. Department of Commerce, in providing the neutron research facilities used in this work. T.S. acknowledges support from the L'Oreal USA Fellowship for Women in Science.
\end{acknowledgments}


\begin{thebibliography}{17}
\expandafter\ifx\csname natexlab\endcsname\relax\def\natexlab#1{#1}\fi
\expandafter\ifx\csname bibnamefont\endcsname\relax
  \def\bibnamefont#1{#1}\fi
\expandafter\ifx\csname bibfnamefont\endcsname\relax
  \def\bibfnamefont#1{#1}\fi
\expandafter\ifx\csname citenamefont\endcsname\relax
  \def\citenamefont#1{#1}\fi
\expandafter\ifx\csname url\endcsname\relax
  \def\url#1{\texttt{#1}}\fi
\expandafter\ifx\csname urlprefix\endcsname\relax\def\urlprefix{URL }\fi
\providecommand{\bibinfo}[2]{#2}
\providecommand{\eprint}[2][]{\url{#2}}

\bibitem[{\citenamefont{Dingle et~al.}(1978)\citenamefont{Dingle, Stormer,
  Gossard, and Wiegmann}}]{Dingle}
\bibinfo{author}{\bibfnamefont{R.}~\bibnamefont{Dingle}},
  \bibinfo{author}{\bibfnamefont{H.}~\bibnamefont{Stormer}},
  \bibinfo{author}{\bibfnamefont{A.}~\bibnamefont{Gossard}}, \bibnamefont{and}
  \bibinfo{author}{\bibfnamefont{W.}~\bibnamefont{Wiegmann}},
  \bibinfo{journal}{Appl.\ Phys.\ Lett.} \textbf{\bibinfo{volume}{33}},
  \bibinfo{pages}{665} (\bibinfo{year}{1978}).

\bibitem[{\citenamefont{Logvenov et~al.}(2009)\citenamefont{Logvenov, Gozar,
  and Bozovic}}]{Bozovic}
\bibinfo{author}{\bibfnamefont{G.}~\bibnamefont{Logvenov}},
  \bibinfo{author}{\bibfnamefont{A.}~\bibnamefont{Gozar}}, \bibnamefont{and}
  \bibinfo{author}{\bibfnamefont{I.}~\bibnamefont{Bozovic}},
  \bibinfo{journal}{Science} \textbf{\bibinfo{volume}{326}},
  \bibinfo{pages}{699} (\bibinfo{year}{2009}).

\bibitem[{\citenamefont{Kozuka et~al.}(2009)\citenamefont{Kozuka, Kim, Bell,
  Kim, Hikita, and Hwang}}]{Kozuka}
\bibinfo{author}{\bibfnamefont{Y.}~\bibnamefont{Kozuka}},
  \bibinfo{author}{\bibfnamefont{M.}~\bibnamefont{Kim}},
  \bibinfo{author}{\bibfnamefont{C.}~\bibnamefont{Bell}},
  \bibinfo{author}{\bibfnamefont{B.~G.} \bibnamefont{Kim}},
  \bibinfo{author}{\bibfnamefont{Y.}~\bibnamefont{Hikita}}, \bibnamefont{and}
  \bibinfo{author}{\bibfnamefont{H.~Y.} \bibnamefont{Hwang}},
  \bibinfo{journal}{Nature} \textbf{\bibinfo{volume}{462}},
  \bibinfo{pages}{487} (\bibinfo{year}{2009}).

\bibitem[{\citenamefont{de~Gennes}(1960)}]{deGennes}
\bibinfo{author}{\bibfnamefont{P.-G.} \bibnamefont{de~Gennes}},
  \bibinfo{journal}{Phys.\ Rev.} \textbf{\bibinfo{volume}{118}},
  \bibinfo{pages}{141} (\bibinfo{year}{1960}).

\bibitem[{\citenamefont{Kawano et~al.}(1996)\citenamefont{Kawano, Kajimoto,
  Kubota, and Yoshizawa}}]{Kawano}
\bibinfo{author}{\bibfnamefont{H.}~\bibnamefont{Kawano}},
  \bibinfo{author}{\bibfnamefont{R.}~\bibnamefont{Kajimoto}},
  \bibinfo{author}{\bibfnamefont{M.}~\bibnamefont{Kubota}}, \bibnamefont{and}
  \bibinfo{author}{\bibfnamefont{H.}~\bibnamefont{Yoshizawa}},
  \bibinfo{journal}{Phys.\ Rev.\ B} \textbf{\bibinfo{volume}{53}},
  \bibinfo{pages}{R14709} (\bibinfo{year}{1996}).

\bibitem[{\citenamefont{Khomskii}(2010)}]{Khomskii}
\bibinfo{author}{\bibfnamefont{D.~I.} \bibnamefont{Khomskii}},
  \emph{\bibinfo{title}{Basic Aspects of the Quantum Theory of Solids}}
  (\bibinfo{publisher}{Cambridge University Press}, \bibinfo{year}{2010}).

\bibitem[{\citenamefont{Dagotto}(2002)}]{Dagotto}
\bibinfo{author}{\bibfnamefont{E.}~\bibnamefont{Dagotto}},
  \emph{\bibinfo{title}{Nanoscale Phase Separation and Colossal
  Magnetoresistance}} (\bibinfo{publisher}{Springer}, \bibinfo{year}{2002}).

\bibitem[{\citenamefont{van~den Brink and Khomskii}(1999)}]{Brink}
\bibinfo{author}{\bibfnamefont{J.}~\bibnamefont{van~den Brink}}
  \bibnamefont{and} \bibinfo{author}{\bibfnamefont{D.}~\bibnamefont{Khomskii}},
  \bibinfo{journal}{Phys.\ Rev.\ Lett.} \textbf{\bibinfo{volume}{82}},
  \bibinfo{pages}{1016} (\bibinfo{year}{1999}).

\bibitem[{\citenamefont{Kuwahara et~al.}(1999)\citenamefont{Kuwahara, Okuda,
  Tomioka, Asamitsu, and Tokura}}]{Kuwahara}
\bibinfo{author}{\bibfnamefont{H.}~\bibnamefont{Kuwahara}},
  \bibinfo{author}{\bibfnamefont{T.}~\bibnamefont{Okuda}},
  \bibinfo{author}{\bibfnamefont{Y.}~\bibnamefont{Tomioka}},
  \bibinfo{author}{\bibfnamefont{A.}~\bibnamefont{Asamitsu}}, \bibnamefont{and}
  \bibinfo{author}{\bibfnamefont{Y.}~\bibnamefont{Tokura}},
  \bibinfo{journal}{Phys.\ Rev.\ Lett.} \textbf{\bibinfo{volume}{82}},
  \bibinfo{pages}{4316} (\bibinfo{year}{1999}).

\bibitem[{\citenamefont{Santos et~al.}(2009)\citenamefont{Santos, May,
  Robertson, and Bhattacharya}}]{Santos_AFmetal}
\bibinfo{author}{\bibfnamefont{T.}~\bibnamefont{Santos}},
  \bibinfo{author}{\bibfnamefont{S.}~\bibnamefont{May}},
  \bibinfo{author}{\bibfnamefont{J.}~\bibnamefont{Robertson}},
  \bibnamefont{and}
  \bibinfo{author}{\bibfnamefont{A.}~\bibnamefont{Bhattacharya}},
  \bibinfo{journal}{Phys.\ Rev.\ B} \textbf{\bibinfo{volume}{80}},
  \bibinfo{pages}{155114} (\bibinfo{year}{2009}).

\bibitem[{sup()}]{suppl_mat}
\bibinfo{note}{See the Supplemental Material}.

\bibitem[{\citenamefont{Fitzsimmons and Majkrzak}(2005)}]{Fitz_chapter}
\bibinfo{author}{\bibfnamefont{M.}~\bibnamefont{Fitzsimmons}} \bibnamefont{and}
  \bibinfo{author}{\bibfnamefont{C.}~\bibnamefont{Majkrzak}},
  \emph{\bibinfo{title}{Modern Techniques for Characterizing Magnetic
  Materials}} (\bibinfo{publisher}{Springer, New York}, \bibinfo{year}{2005}),
  chap.~\bibinfo{chapter}{3}, pp. \bibinfo{pages}{107--155}.

\bibitem[{not()}]{note_qvalues}
\bibinfo{note}{The SL periodicity (9$c$) is found from the distance between the
  two peaks in Fig.~3a: $0.82$\AA$^{-1} - 0.64$\AA$^{-1} = 0.18$\AA$^{-1} =
  2\pi/(9c)$.}

\bibitem[{\citenamefont{Chmaissem et~al.}(2003)\citenamefont{Chmaissem,
  Dabrowski, Kolesnik, Mais, Jorgensen, and Short}}]{Chmaissem}
\bibinfo{author}{\bibfnamefont{O.}~\bibnamefont{Chmaissem}},
  \bibinfo{author}{\bibfnamefont{B.}~\bibnamefont{Dabrowski}},
  \bibinfo{author}{\bibfnamefont{S.}~\bibnamefont{Kolesnik}},
  \bibinfo{author}{\bibfnamefont{J.}~\bibnamefont{Mais}},
  \bibinfo{author}{\bibfnamefont{J.~D.} \bibnamefont{Jorgensen}},
  \bibnamefont{and} \bibinfo{author}{\bibfnamefont{S.}~\bibnamefont{Short}},
  \bibinfo{journal}{Phys.\ Rev.\ B} \textbf{\bibinfo{volume}{67}},
  \bibinfo{pages}{094431} (\bibinfo{year}{2003}).

\bibitem[{\citenamefont{Kiyama et~al.}(2003)\citenamefont{Kiyama, Wakabayashi,
  Nakao, Ohsumi, Murakami, Izumi, Kawasaki, and Tokura}}]{Kiyama}
\bibinfo{author}{\bibfnamefont{T.}~\bibnamefont{Kiyama}},
  \bibinfo{author}{\bibfnamefont{Y.}~\bibnamefont{Wakabayashi}},
  \bibinfo{author}{\bibfnamefont{H.}~\bibnamefont{Nakao}},
  \bibinfo{author}{\bibfnamefont{H.}~\bibnamefont{Ohsumi}},
  \bibinfo{author}{\bibfnamefont{Y.}~\bibnamefont{Murakami}},
  \bibinfo{author}{\bibfnamefont{M.}~\bibnamefont{Izumi}},
  \bibinfo{author}{\bibfnamefont{M.}~\bibnamefont{Kawasaki}}, \bibnamefont{and}
  \bibinfo{author}{\bibfnamefont{Y.}~\bibnamefont{Tokura}},
  \bibinfo{journal}{J.\ Phys.\ Soc.\ J.} \textbf{\bibinfo{volume}{72}},
  \bibinfo{pages}{785} (\bibinfo{year}{2003}).

\bibitem[{\citenamefont{Lin et~al.}(2006)\citenamefont{Lin, Okamoto, and
  Millis}}]{Millis}
\bibinfo{author}{\bibfnamefont{C.}~\bibnamefont{Lin}},
  \bibinfo{author}{\bibfnamefont{S.}~\bibnamefont{Okamoto}}, \bibnamefont{and}
  \bibinfo{author}{\bibfnamefont{A.}~\bibnamefont{Millis}},
  \bibinfo{journal}{Phys.\ Rev.\ B} \textbf{\bibinfo{volume}{73}},
  \bibinfo{pages}{041104(R)} (\bibinfo{year}{2006}).

\bibitem[{\citenamefont{May et~al.}(2008)\citenamefont{May, Shah, te~Velthuis,
  Fitzsimmons, Zuo, Zhai, Eckstein, Bader, and Bhattacharya}}]{May_PRB07}
\bibinfo{author}{\bibfnamefont{S.~J.} \bibnamefont{May}},
  \bibinfo{author}{\bibfnamefont{A.~B.} \bibnamefont{Shah}},
  \bibinfo{author}{\bibfnamefont{S.~G.~E.} \bibnamefont{te~Velthuis}},
  \bibinfo{author}{\bibfnamefont{M.~R.} \bibnamefont{Fitzsimmons}},
  \bibinfo{author}{\bibfnamefont{J.~M.} \bibnamefont{Zuo}},
  \bibinfo{author}{\bibfnamefont{X.}~\bibnamefont{Zhai}},
  \bibinfo{author}{\bibfnamefont{J.~N.} \bibnamefont{Eckstein}},
  \bibinfo{author}{\bibfnamefont{S.~D.} \bibnamefont{Bader}}, \bibnamefont{and}
  \bibinfo{author}{\bibfnamefont{A.}~\bibnamefont{Bhattacharya}},
  \bibinfo{journal}{Phys.\ Rev.\ B} \textbf{\bibinfo{volume}{77}},
  \bibinfo{pages}{174409} (\bibinfo{year}{2008}).

\end{thebibliography}
\end{document}